\documentclass[USletter,9pt,twocolumn,twoside]{article}

\usepackage[utf8]{inputenc}
\usepackage[english]{babel}
\usepackage[T1]{fontenc}
\usepackage{url}
\usepackage[colorlinks=true, allcolors=blue]{hyperref}
\usepackage{graphicx}
\usepackage[left=45pt, right=45pt, top=50pt, bottom=65pt,
headheight=15pt,%
headsep=10pt]{geometry}%
\usepackage{amsmath}
\usepackage{authblk}
\usepackage{algorithm, algpseudocode}
\usepackage{mathptmx}

\usepackage[figurename=Fig.]{caption}
                
\renewenvironment{abstract}
{
\noindent \rule{\linewidth}{.5pt}\medskip\par{\bfseries \abstractname.}}
{\noindent \rule{\linewidth}{.5pt}

}

\title{Holographic beam shaping of partially coherent light}

\author[1,*]{Nicolas Barré}
\author[1]{Alexander Jesacher}

\affil[1]{Institute of Biomedical Physics, Medical University of
  Innsbruck, M\"{u}llerstra\ss e 44, 6020 Innsbruck, AT}

\affil[*]{Corresponding author: nicolas.barre@i-med.ac.at}

\begin{document}

\maketitle

\begin{abstract}
  We present an algorithm for holographic shaping of partially
  coherent light, bridging the gap between traditional coherent and
  geometric optical approaches. The description of partially coherent
  light relies on a mode expansion formalism, with possibly thousands
  of individual modes, and the inverse-design optimization algorithm
  is based on gradient descent with the help of algorithmic
  differentiation. We demonstrate numerically and experimentally that
  an optical system consisting of two phase patterns can potentially
  achieve any intensity profile transformation with good accuracy.
\end{abstract}

Sculpting light freely into user-defined shapes is a central key
aspect in many areas of photonics. The list of fields with a strong
interest in beam shaping is extensive and ranges from applied areas
such as laser material processing~\cite{hafner2018tailored,
  flamm2019beam, salter2019adaptive, shi2020microstructural}, mode
conversion~\cite{fontaine2019laguerre}, near-eye
displays~\cite{chakravarthula2019wirtinger}, lithography,
telecommunications and practical lighting systems to fundamental
research in biology and physics, such as investigations towards
revealing the functioning of the brain via patterned
photostimulation~\cite{ronzitti2017recent} or the trapping of
microparticles~\cite{otte2020optical} and
atoms~\cite{barredo2018synthetic}.  Complex spatial wavefronts can be
shaped using either static components such as diffractive optical
elements (DOEs), freeform elements~\cite{Ries:02} and
metasurfaces~\cite{kamali2018review}, or dynamic devices like spatial
light modulators (SLM) based on liquid crystal, micro
electromechanical systems (MEMS) technology, or even
soundwaves~\cite{bechtold2013beam}.

Despite the prowess of today's beam shaping technology, there remain
important challenges. One of them is tailoring fields with a lack of
coherence, i.e. partially coherent or incoherent light, which is of
central importance in many lighting and high power laser
applications. 
In particular configurations, where light sources satisfy the
geometric optical regime, some solutions based on ray-optical
approaches have proven to be efficient~\cite{Oliker:18,Wei:19}.  The
main advantage of such approaches is that they enforce the design of
surfaces with curl-free gradients that are perfectly adapted to
freeform elements.  However, these methods inherently neglect
diffraction and are therefore ill-suited for applications involving
miniaturized integrated optics~\cite{Schmidt:20} or multicolor beam
shaping~\cite{jesacher2014colour, wang2018dynamic,
  chakravarthula2019wirtinger}, as well as for applications requiring
more complex partially coherent sources.

Usual representations of partially coherent light introduce
4-dimensional second-order correlation
functions~\cite{born2013principles}, like the mutual coherence
function or the mutual intensity function in the quasi-monochromatic
case, in order to accurately describe the statistical nature of light
fields seen as random processes. Such 4D correlation functions admit,
due to Mercer's theorem, a mode expansion consisting of spatially
coherent and mutually uncorrelated fields. Although it is also
admitted that computing such mode expansions is in general a hard
task, as it involves solving the eigenvalues of a Fredholm integral
equation, some ubiquitous light sources like multimode lasers or
multimode fibers already exhibit eigen-modes as the basis of their
physical description. Moreover, for different types of sources, it is
often possible to find some phenomenological approximate models that
provide convenient analytical modal descriptions, for instance in the
same way as we proceed in the following to model a light-emitting
diode (LED) source.

In this article we present a novel algorithmic approach, which allows
the efficient design of multiple phase-only DOEs for shaping user
defined intensity distributions of partially coherent light, as
described above in terms of spatially coherent and mutually
uncorrelated modes. The approach employs numerical wave propagation
and is not limited by any geometric constraints such as rotational
symmetry. Its only limitation lies in the thin element approximation
(TEA) which must hold for the experimental realization.  For the sake
of generality, we provide an algorithm that accounts for an arbitrary
number of DOEs, although we show later that two phase patterns are
sufficient in all the situations we considered, to obtain nearly
perfect intensity shaping. Moreover, we experimentally demonstrate the
concept by redistributing a part of the light from an LED into various
shapes using two diffractive patterns displayed on a liquid crystal
SLM. Given the widespread use of partially coherent sources in science
and technology, we anticipate that our work will be helpful for the
advancement of research fields concerned with tailored light fields.

The algorithm we developed consists of an optimization procedure
relying on gradient descent in order to minimize the $l_2$ distance
between the time-averaged transverse intensity profile $I_S$ of the
partially coherent source after crossing the optical system and a
user-defined target intensity profile $I_T$. Since we assume that the
modes defining the partially coherent source are mutually
uncorrelated, their time-averaged intensity $I_S$ is simply the sum of
the contributions of all the individual modes constituting the
source. The optimization procedure is iterative and consists of three
stages. First, a forward pass propagates $n_{modes}$ input modes
$U_0[1:n_{modes}]$ from the partially coherent source plane to the
detection plane where the desired intensity profile $I_T$ is to be
detected, through the optical system consisting of $n_{layers}$
longitudinally separated phase patterns $\Phi[1:n_{layers}]$. A
pseudo-code description of the forward model is presented in
algorithm~\ref{alg:forward_model}. Second, the time-averaged input
intensity $I_S$ is computed in the detection plane and serves to
define an error value $err$ based on the $l_2$ distance between $I_S$
and $I_T$. Since the goal of the algorithm is to minimize this error
with respect to the phase parameters $\Phi$, the gradient
$\nabla_\Phi$ of this error has to be computed. To this end, we
develop a backward pass, as is commonly done in computer science to
numerically compute the gradients of complicated neural network
architectures with gradient backpropagation algorithms. Here we used a
few rules from the set defined in~\cite{Jurling:14}, leading to the
error and gradient computation
algorithm~\ref{alg:gradient_computation}. Third, we need to update the
phase patterns $\Phi$. Since we managed to compute a gradient, any
first-order optimization method can be employed, the simplest being
adding a small gradient step of opposite direction,
$\Phi \gets \Phi - \alpha\nabla_\Phi$ with $\alpha > 0$. These three
stages need to be repeated iteratively, until a convergence criterion
regarding the error is achieved, $err < \epsilon$ with $\epsilon > 0$.

\begin{algorithm}
\caption{Forward model}\label{alg:forward_model}
\begin{algorithmic}
\State $\mathbf{Inputs:}$ $U_0$ a list of 2D input modes\\
\phantom{$\mathbf{Inputs:}$} $\Phi$ a list of phase patterns
\State $\mathbf{Outputs:}$ $U$ a list of propagated modes\\
\phantom{$\mathbf{Outputs:}$} $S$ a list of modes stored in each plane
\Procedure{propagate\_forward}{$U_0$, $\Phi$}
\State $U \gets \Call{copy}{U_0}$ \Comment{copy initialization}
\State $S \gets \Call{Modes\_List}{n_{layers}}$
\Comment empty initialization
\For{$l=1\,\mathbf{to}\,n_{layers}$}
  \State $U \gets \Call{prop.}{U, l}$ \Comment{propagating modes to plane $l$}
  \State $U \gets U\cdot\exp{(i \Phi[l])}$
  \Comment{crossing $l$\textsuperscript{th} DOE forward}
  \State $S[l] \gets U$ \Comment{storing modes at plane $l$}
\EndFor
\State $U \gets$ \Call{FFT.}{$U$} \Comment{propagating to detection plane}
\State \Return $(U,S)$
\EndProcedure
\end{algorithmic}
\end{algorithm}

In algorithm~\ref{alg:forward_model}, the inputs of the
\textsc{propagate\_forward} procedure are $U_0$, a list of $n_{modes}$
transverse modes mutually uncorrelated constituting the partially
coherent source, and $\Phi$ a list of $n_{layers}$ phase patterns
initially set to flat phases. The outputs are $U$, a list of modes
propagated to the detection plane, and $S$, a list of modes stored
just after crossing each phase pattern which will be essential to
compute the backpropagated gradient later on. More precisely, $S$ is a
list of a list, the outer list being of length $n_{layers}$ and
storing inner lists of modes of length $n_{modes}$. The data
structures present in the algorithms are defined in order to simplify
the notations and provide conciseness, but do not prejudge how they
should be implemented in practice in order to get the best
performance. In order to avoid verbose loops on the mode index, we
present the code in a vectorized form by appending a \emph{dot} ($.$)
after the functions that we want to broadcast over the different
elements of a list type parameter. Similarly, the vectorized
point-wise multiplication of a collection of modes with a single phase
pattern is defined by a \emph{dot} operator ($\cdot$). To describe
mode propagation from its current plane definition to a phase pattern
plane of index $l$, we introduce a generic \textsc{prop} function,
taking both a tranverse mode and an index $l$ as parameters, which can
be defined according to the user's choice. A simple model that we also
use in the current study is the Angular Spectrum (AS)
method 
for free-space propagation and Fast Fourier Transform (\textsc{FFT})
for Fourier collimation or focusing. In the forward model
description~\ref{alg:forward_model} as well as in our following
experimental demonstration, we choose the detection plane to be
Fourier-conjugated to the last phase pattern of the optical system,
but this is again arbitrary, up to the user's choice.

\begin{algorithm}
\caption{Error and gradient computation}\label{alg:gradient_computation}
\begin{algorithmic}
\State $\mathbf{Inputs:}$ $U$ a list of propagated modes\\
\phantom{$\mathbf{Inputs:}$} $S$ a list of modes stored in each plane\\
\phantom{$\mathbf{Inputs:}$} $\Phi$ a list of phase patterns\\
\phantom{$\mathbf{Inputs:}$} $I_T$ a target intensity profile
\State $\mathbf{Outputs:}$ $err$ an error associated with the forward model
\phantom{$\mathbf{Outputs:}$} $\nabla_\Phi$ a list of phase gradients
\Procedure{backpropagate\_gradient}{$U$, $S$, $\Phi$, $I_T$}
\State $I_S \gets \sum\limits_{m=1}^{n_{modes}}\vert U[m]\vert^2$
\Comment storing source intensity
\State $err \gets {\displaystyle \int}{\left(I_S - I_T\right)^2}$
\Comment computing the error value
\State $\nabla_\Phi \gets \Call{Phase\_List}{n_{layers}}$
\Comment empty initialization
\State $U \gets 4\left(I_S - I_T\right)\cdot U$
\State $U \gets \Call{IFFT.}{U}$ \Comment{backpropagating to the last DOE}
\For{$l=n_{layers}\,\mathbf{to}\,1$}
\State $\nabla_\Phi[l] \gets \Im{\left(\sum\limits_{m=1}^{n_{modes}}(S^*[l]\cdot U)[m]\right)}$ \Comment{$l$\textsuperscript{th} phase gradient}
\State $U \gets U\cdot\exp{(-i \Phi[l])}$
\Comment{crossing $l$\textsuperscript{th} DOE backward}
\State $U \gets \Call{prop.*}{U, l}$ \Comment{propagating backward}
\EndFor
\EndProcedure
\State \Return $(err,\nabla_\Phi)$
\end{algorithmic}
\end{algorithm}

The algorithm~\ref{alg:gradient_computation} describes the error and
gradients computation which are returned by the
\textsc{backpropagate\_gradient} procedure. The list of propagated
fields $U$ returned by the forward propagation procedure is used to
compute the error $err$ and is backpropagated in a mirrored fashion
through the same optical system. In order to achieve gradient
backpropagation, all the propagation operators need to be replaced by
their adjoint (Hermitian conjugate) operators denoted by an asterisk
($^{*}$). Moreover, it is important to notice that the gradient
computation $\nabla_\Phi[l]$ associated to each individual phase
pattern $\Phi[l]$ occurs in reverse order and depends only on the
modes stored in $S[l]$ during the forward pass and on the modes $U$
backpropagated to the same plane $l$.

\begin{figure}[t]
    \centering
    \includegraphics{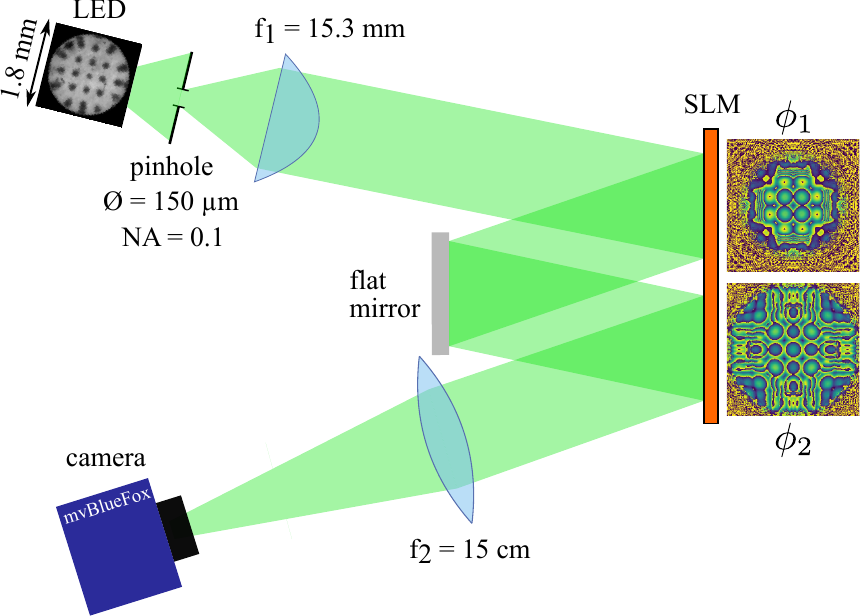}
    \caption{Folded setup for holographic beam shaping with 2 reflections on a single SLM.}
    \label{fig:setup}
\end{figure}

We experimentally demonstrate the performance and accuracy of our beam
shaping algorithm by implementing the setup shown in
Fig.~\ref{fig:setup}.  A single SLM (Hamamatsu X10468, $16$~mm
$\times$ $12$~mm) and a mirror (square with 5~mm side length) form a
tandem configuration of two diffractive patterns. Similar layouts have
been used in the past for shaping coherent light and approaching
properties of volume holograms, such as wavelength and angular
selectivity~\cite{Bartelt:85, Borgsmuller:03, Jesacher:08,
  wang2018dynamic, scholes2020improving, MoserPsaltis2020}.  Our
approach is conceptually related, however, our algorithm is different
and the number of modes it manages to handle in order to accurately
describe partially coherent light is substantially higher compared to
the previous state of the art (up to 1600 individual coherent modes
shown in this letter), which is an essential prerequisite for
realizing practical application cases.

In this proof of concept study, we employ a green LED (Thorlabs
M530L4) as a primary source, with a peak emission wavelength around
$522$~nm and $35$~nm bandwidth. The emitter is square-shaped with 1~mm
side length with a plastic dome affixed on the top that modifies the
spatial emission characteristics.
Following a pragmatic approach, we incorporate the optical effects of
the plastic dome by modelling the LED according to its image through a
1:1 Keplerian telescope (NA = 0.25), where the emitter appears to be
roughly circular with a radius of $r_0 = 900$~\textmu m, exhibiting
several dark spots arranged in a square pattern. In our algorithm, we
use this image as a replacement for the true LED plus the dome lens.
The LED light is filtered by a pinhole of radius
$r_{pin} = 75$~\textmu m, which is positioned at a distance $z_{pin}$
from the LED such that the numerical aperture (NA) of the light
passing through it is about $\textrm{NA} \simeq 0.1$. After
collimation by a first lens (focal length $f_1$ = 15.3~mm), the beam
radius is measured to be $r_{col} = 1625$~\textmu m.
After collimation, the light traverses two diffractive patterns,
$\Phi_1$ and $\Phi_2$, at distances of 165~mm and 325~mm from the
collimating lens, respectively. Finally, it is directed onto a camera
by a second lens ($f_2$ = 15~cm) at a distance $f_2$ from the
sensor. Therefore, the camera captures a magnified image of the
aperture ($\approx\times 10$) if no diffractive patterns are shown.


From the simulation point of view, we treat the LED as a totally
incoherent light source~\cite{deng2017:coherence}. We discretize the
LED emitter model into a square grid of size $N_S\times N_S$, which
determines the number of uncorrelated modes representing our
source. The pixel size $dx_S = dy_S$ in each dimension is chosen such
that the total grid side length is $2r_0$, and the source profile is
idealized to a disc of radius $r_0$ with constant intensity. Then,
each pixel acts as an independent spherical wave source with an
amplitude proportional to the square root of the local intensity,
$\sqrt{I}e^{i k r}/r$, that propagates to the pinhole plane where it
is clipped by the 150~\textmu m diameter aperture. Thus, the
analytical expression of each transverse mode is known in the pinhole
plane and serves to define $N_S^2$ input distributions stored in grids
of size $512\times512$ with a pixel size
$dx_{pin} = dy_{pin} = 1$~\textmu m. The collimation of the input
modes by the lens following the pinhole is modeled via a scaled FFT
(Chirp Z-transform), and the resolution after collimation is set to
$dx = dy = 20$~\textmu m, which directly matches the SLM pixel
size. Obviously, given our disc shape model of the LED, some of the
generated modes have zero intensity and could be removed for
optimizing memory and computation time (the number of effectively used
modes is about $\pi N_S^2/4$).

\begin{figure}[t]
    \centering
    \includegraphics{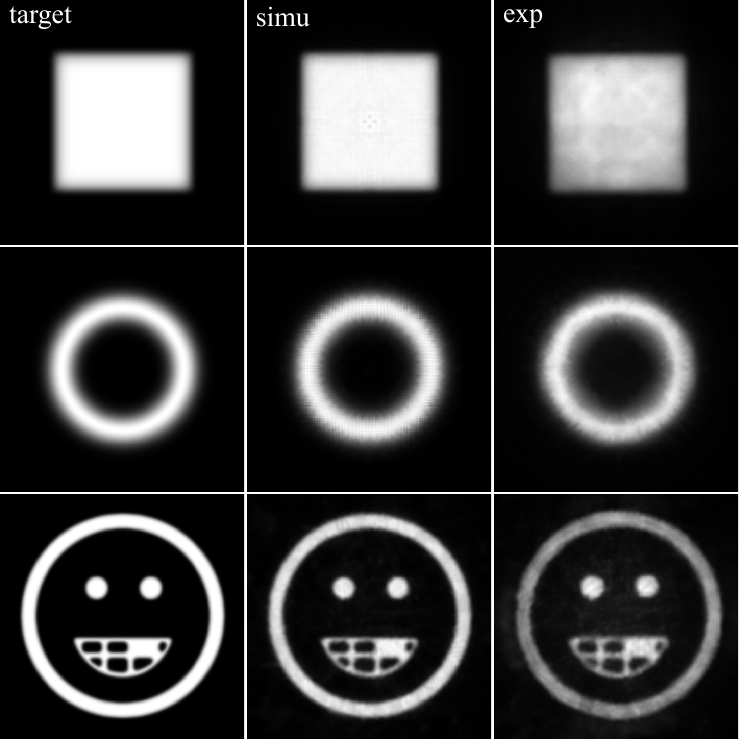}
    \caption{Results of the beam shaping algorithm with 2 phase
      patterns for 3 different target intensity profiles (square,
      ring, smiley). The first column represents the target intensity
      profiles set as inputs of the simulation. The second column
      shows the simulated intensity profiles and the third column
      shows the intensity profiles recorded on a camera when the phase
      patterns computed by the simulation are displayed on the SLM.}
    \label{fig:2p_shapes}
\end{figure}

The results of the beam shaping algorithm with two phase patterns are
displayed in Fig.~\ref{fig:2p_shapes}. We chose $N_S=40$ corresponding
to 1600 modes. In order to demonstrate the flexibility of the
algorithm, the LED emission is sculpted into three different shapes: a
square of 1.5~mm side length, a Gaussian ring with $675$~\textmu m
radius and $150$~\textmu m waist, and a smiley containing fine details
with 2.3~mm outer diameter. The simulations show an excellent
agreement with the theoretical target intensity profiles, which are
obtained after a few hundreds of iterations associated to an error
bound $\epsilon = 1.10^{-8}$, for targets of total intensity
normalized to unity. The typical computation time for such precision
is about 10 minutes on a mid-range GPU (Geforce RTX 2070 Super). The
experimental results obtained with setup~\ref{fig:setup} are likewise
excellently matching the theoretical targets, even if we sometimes
observe some weak modulations reminiscent of the LED substructure that
we did not fully take into account.  Moreover, we experimentally
observe that our multimode holographic transformations are much less
sensitive to misalignment than similar shaping of coherent light,
where noticeable speckle usually appears even for small misalignments.

\begin{figure}[t]
    \centering
    \includegraphics{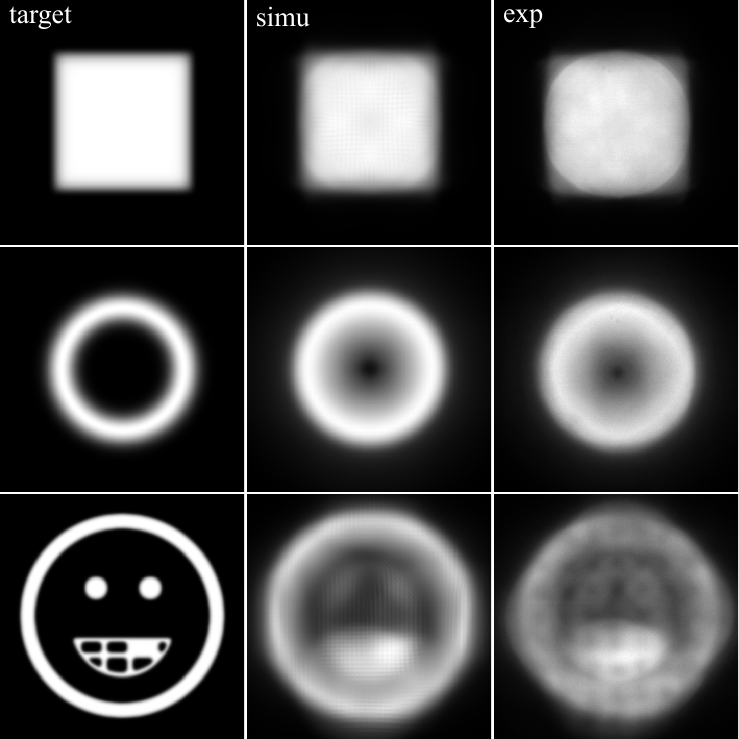}
    \caption{Results of the beam shaping algorithm with 1 phase
      pattern for the same targets as in figure~\ref{fig:2p_shapes}.}
    \label{fig:1p_shapes}
\end{figure}

Our simulations and experiments demonstrate that two phase patterns
are sufficient to obtain nearly perfect intensity shaping of an
incoherent source such as an LED. We performed further simulations
considering different mode bases suitable for describing several other
partially incoherent light sources, such as Hermite-Gaussian (HG) or
Linearly Polarized (LP) modes, obtaining similar quality levels for
all cases (data not shown).

Next we show that these complicated beam transformations cannot be
achieved by a single phase pattern. For this purpose we enforce
$\Phi_1$ to remain flat, thus only relying on $\Phi_2$ to achieve the
same transformations as above. The results displayed in
Fig.~\ref{fig:1p_shapes} confirm that one phase pattern is not
sufficient to ensure a good convergence of the algorithm to the
defined targets. Notably, we still observe a good agreement between
simulations and experimental results.

\begin{figure}
    \centering
    \includegraphics{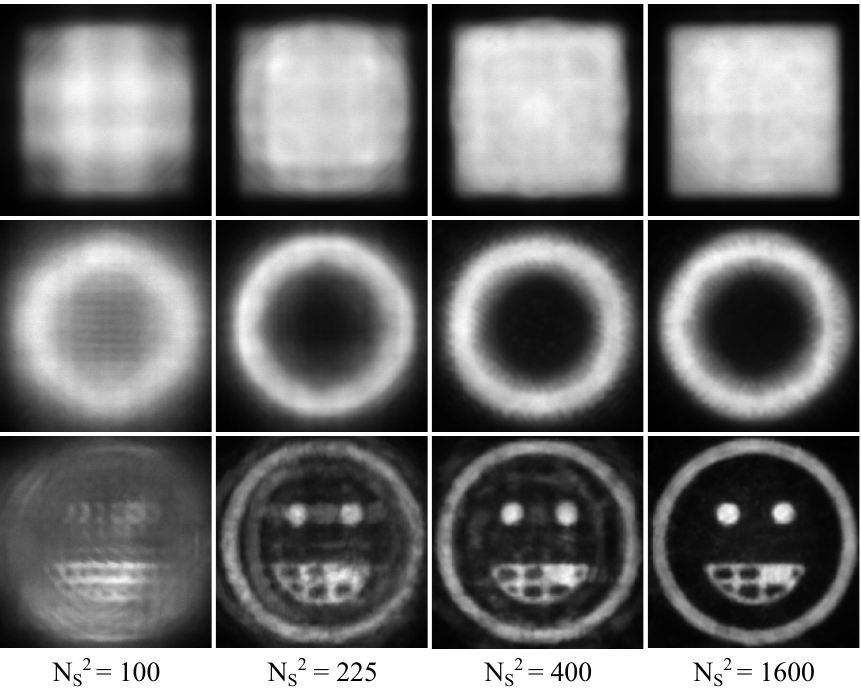}
    \caption{Evolution of the experimental intensity profiles with
      respect to the number of modes involved in the modeling of the
      input beam.}
    \label{fig:modes_evolution}
\end{figure}

Finally, it is important to discuss the impact of the number of modes
used for source modeling on the reconstruction quality. Regardless the
number of modes, the simulation always converges almost perfectly to
the given target intensity, but severely undercutting the number of
modes required to accurately describe the light source significantly
degrades the result. Figure~\ref{fig:modes_evolution} visualizes the
influence of the mode number on the quality of the experimental
intensity shaping. An upper bound for the number of modes required to
represent our light source can be estimated using Shannon's sampling
theorem. The maximally allowed spatial sampling distance for any
object imaged at a given NA is $dx_{min} = \lambda/(2\textrm{NA})$. We
can determine the maximum number of modes to be
$n_{max} = \pi r_{pin}^2/dx_{min}^2$, which is close to
$n_{max} = 2600$ in our case. We managed to obtain a very good
intensity fidelity with only about 1257 modes for our three profiles,
but finer details or sharper edges in the target intensity profile may
require an even higher number.

In conclusion, we presented a general algorithm designed to shape the
intensity of partially coherent light, given its expansion in terms of
coherent and mutually uncorrelated modes. We demonstrated the
efficiency of this algorithm and evaluated it experimentally with a
simple optical system consisting of two phase-only diffractive
patterns, which appears sufficient to achieve a very good fidelity.
Incorporating 1600 modes as presented in this work takes about 10
minutes of computation time. From this we can conclude that handling
light sources of substantially higher complexity (e.g. 10$\times$ more
modes) would readily work in a reasonable time on standard PC
hardware.  Therefore, we can envision extending this work to more
complex situations such as the combination of several partially
coherent beams, eventually with multiple colors, by adapting this
algorithm to the design of multilayer or volumic DOEs.

\paragraph{Funding.}
The work is funded by the FWF (I3984-N36).

\clearpage

\let\OLDthebibliography\thebibliography
\renewcommand\thebibliography[1]{
  \OLDthebibliography{#1}
  \setlength{\parskip}{1ex}
  \setlength{\itemsep}{0pt plus 1ex}
}

\bibliographystyle{unsrt}
\bibliography{intensity_shaping}

\end{document}